\documentclass[12pt]{article}
\usepackage{epsfig}
\usepackage{tabularx,pifont,multirow}

\def\beq{\begin{equation}}
\def\eeq#1{\label{#1}\end{equation}}
\def\eeqn{\end{equation}}

\def\beqa{\begin{eqnarray}}
\def\eeqa#1{\label{#1}\end{eqnarray}}
\def\eeqan{\end{eqnarray}}

\let\bar=\overbar

\def\Dslash{\not{\hbox{\kern-4pt $D$}}}
\def\dslash{\not{\hbox{\kern-2pt $\del$}}}

\def\BR{\mbox{\rm BR}}
\def\ee{e^+e^-}

\def\msb{{\bar{\ssstyle M \kern -1pt S}}}

\def\s#1{\widetilde{#1}}

\input babarsym.tex
\newcommand{\gevmone}{\ensuremath{\mathrm{\,Ge\kern -0.1em V^{-1}}}\xspace}

\newcommand{\bei}{\begin{itemize}}
\newcommand{\eei}{\end{itemize}}
\newcommand{\beqns}{\begin{eqnarray*}}
\newcommand{\eeqns}{\end{eqnarray*}}

\def\lnL{\ln{\cal L}}

\def\calC{{\ensuremath{\cal C}}\xspace}
\def\KS{\ensuremath{K^0_S}}

\def\fscfave{\kern 0.18em\overline{\kern -0.18em f}_{\rm SCF}}

\def\Abar{\kern 0.18em\overline{\kern -0.18em A}{}}

\def\Amptpbar{\kern 0.18em\overline{\kern -0.18em {\cal A}}}

\newcommand{\Ksksks}             {\mbox{$\KS\KS\KS$}}

\def\pipi   {\ensuremath{\pip\pim}\xspace}

\newcommand{\fI}                 {\mbox{$f_0(980)$}}

\def\betaeff{\beta_{\rm eff}}

\def\be{\begin{equation}}
\def\ee{\end{equation}}
\def\bea{\begin{eqnarray}}
\def\eea{\end{eqnarray}}

\def\DDstarm {\ensuremath{D^{(*)-}}}
\def\DDstarp {\ensuremath{D^{(*)+}}}

\newcommand{\bkkkboth}{\ensuremath{\Bp \rightarrow \Kp\Km\Kp}\xspace}

\newcommand{\bkkks}{\ensuremath{\Bz \rightarrow \Kp\Km\KS}\xspace}

\newcommand{\kkks}{\ensuremath{\Kp\Km\KS}\xspace}

\newcommand{\bkksks}{\ensuremath{\Bp \rightarrow \KS\KS\Kp}\xspace}

\newcommand{\bkkkgeneric}{\ensuremath{\B\rightarrow\kaon\kaon\kaon}\xspace}

\newcommand{\Kspp}{\ensuremath{\KS\to\pip\pim}\xspace}
\newcommand{\Kszz}{\ensuremath{\KS\to\piz\piz}\xspace}

\newcommand{\Acp}{\ensuremath{A_{\CP}}\xspace}

\newcommand{\phiI}{\ensuremath{\phi(1020)}\xspace}

\newcommand{\ModeO}{\ensuremath{K\pi}}
\newcommand{\ModeT}{\ensuremath{K\pi\piz}}
\newcommand{\ModeTh}{\ensuremath{K\pi\pi\pi}}
\newcommand{\ModeF}{\ensuremath{\KS\pi\pi}}

\def\mrec{\ensuremath{m_{\rm rec}}\xspace}

\def\Sp{\ensuremath{S_+}\xspace}

\def\Cp{\ensuremath{C_+}\xspace}

\renewcommand{\pipi}{\ensuremath{\pi\pi}\xspace}

\def\Title#1{\begin{center} {\Large {\bf #1} } \end{center}}

\begin{document}

\Title{Results on $\beta$ from \babar}

\bigskip\bigskip

%+\addtocontents{toc}{{\it D. Reggiano}}
%+\label{ReggianoStart}

\begin{raggedright}  
{\it Eli Ben-Haim\index{Ben-Haim, E.}\\
Laboratoire de Physique Nucl\'eaire et de Hautes Energies\\
CNRS-IN2P3, Universit\'e Pierre et Marie Curie, Universit\'e Paris Diderot\\
4 place Jussieu, F-75005 Paris, FRANCE\\[0.3cm]
On behalf of the \babar\ collaboration}
\bigskip\bigskip
\end{raggedright}

\noindent
{\it Proceedings of CKM 2012, the 7th International Workshop on the CKM Unitarity Triangle, University of Cincinnati, USA, 28 September - 2 October 2012}

\section{Introduction}

The CKM paradigm for quark mixing in the standard model accounts for the observed \CP-symmetry violation in the quark sector. The CKM mechanism describes all transitions between quarks in terms of only four parameters: three rotation angles and one irreducible phase.
Consequently, the flavor sector of the standard model
is highly predictive.

One particularly interesting prediction is that
the direct and mixing-induced \CP asymmetries, denoted \calC and \calS,
respectively, are approximately the same in decays governed by $b \to
q\bar{q}s$ ($q = u,d,s$) transitions and those found in $b \to c\bar{c}s$ transitions. The level of approximation was estimated as~$\sim 1\%$~\cite{Cheng:2005ug}.
The dominant contributions to $b \to c\bar{c}s$ transitions (typically using $B^0 \to J/\psi \KS$) are from tree-level diagrams.
On the other hand, since flavor changing neutral currents are forbidden at
tree level in the standard model, the $b \to q\bar{q}s$  transition proceeds
almost uniquely via loop diagrams (penguins). In these channels, the weak phases in the respective dominant contributions are the same as in the standard model, which explains the prediction concerning the approximate equality between the \CP-violation parameters.

The CKM angle $\beta$ is related to the mixing-induced \CP\ asymmetry by $\calS = \sin(2\beta)$. Yet, other
amplitudes than penguin may contribute, to some extent, to $b \to q\bar{q}s$ transitions (e.g., ``tree pollution''). The angle extracted from the corresponding \calS parameter is therefore denoted $\betaeff$ rather than $\beta$. Penguin diagrams are affected by new particles in many extensions of the standard model. 
As there may be \CP-violating sources beyond the standard model, these penguin-dominated charmless hadronic $B$ decays, are of great interest because they are sensitive to new physics effects at large energy scales.

Various $b \to s$ dominated charmless hadronic $B$
decays have been studied in order to probe this prediction.
by comparing the mixing-induced \CP asymmetry measured for each mode to that measured in $b \to c\bar{c}s$ transitions.

%Compilations of results~\cite{Amhis:2012bh} show that most of them have central values below that for $b \to c\bar{c}s$.

Here we present recent time-dependent analyses of \Bz meson decays to \Ksksks\ and $\Kp\Km\Kz$ from the \babar\ collaboration. The former is a \CP eigenstate, while in the latter the \CP content is determined by means of an amplitude analysis. They are described in detail in Refs.~\cite{Lees:2011nf} and~\cite{Lees:2012kxa}. They both benefit from small theoretical uncertainties~\cite{Cheng:2005bg}, which make the comparison with $\b\to\ccbar\s$ transitions more meaningful. 

%Recent theoretical
%evaluations~\cite{Grossman:2003qp,Gronau:2003kx,Gronau:2004hp,Cheng:2005bg,Gronau:2005gz,Beneke:2005pu,Engelhard:2005hu,Cheng:2005ug,Williamson:2006hb}
%suggest that standard-model corrections to the $b \to q\bar{q}s$ mixing-induced
%\CP violation parameters should be small, in particular for the modes
%$\phi K^0$, $\eta^\prime K^0$, and $\KS\KS\KS$ (the latter is studied in this work).
%The uncertainties in these predictions are of the order of $0.05$ or even smaller for $\phi\KS$, $\etapr\KS$ and $\Ksksks$.

Measurement of $b\rightarrow c\cbar d$ transitions such as \Bz\to\DDstarp\DDstarm\
should also yield the same value of $\sin(2\beta)$ as in $b \to c\bar{c}s$, to the extent that the contributions
from penguin processes may be neglected. The leading and sub-leading order processes contributing to \Bz\to\DDstarp\DDstarm‏\ decays are described by a tree diagram and a penguin diagram, respectively.
The effect of neglecting the penguin amplitude has been estimated in models 
based on factorization and heavy quark symmetry, and the corrections 
are found to be a few percent\,\cite{Xing:1998ca,Xing:1999yx}.
Loops involving non-SM particles (for example, charged Higgs or SUSY particles) could increase the contribution from penguin diagrams and introduce additional phases. A large deviation of the measured \CP asymmetry \calS from the value of \stwob measured in $b\rightarrow c\cbar s$ transitions
or a non-zero value of direct \CP violation
%\,\cite{Grossman:1996ke,Gronau:2008ed,Zwicky:2007vv}
would be strong evidence of new physics.
Here we present the latest measurement of time-dependent \CP asymmetry of partially reconstructed \Bztodstdst\ decays from the \babar\ collaboration~\cite{Lees:2012px}.

The three analyses below use the final \babar\ \FourS\ dataset that consists of $468\times10^{6}$ \BB\ decays.

%%%%%%%%%%%%%%%%%%%%%%%%%%%%%%%%%%%%%%%%%%%%%%%%%%%%%%%%
\section{Time-dependent \CP asymmetry of $B^0 \to K_S^0 K_S^0 K_S^0$ decays}

The time-dependent \CP-violation parameters ${\cal S}$ and ${\cal C}$ are extracted by modeling the proper-time distribution of $\Bz\to\KS\KS\KS$ decays with $\KS\to\pip\pim$,
%denoted by \pippim, 
and events where one of the \KS\ mesons decays to $\piz\piz$.
%, denoted by \pizpiz.
%
%The maximum-likelihood fit of $3261$ candidates in the $\pippim$ submode and $7209$ candidates in the $\pizpiz$ submode results in the event yields detailed in Table~\ref{tab:td_yields}.
%
%\begin{table}[htb]
%\begin{center}
%\caption{ \label{tab:td_yields} Event yields for the different event species, resulting from the maximum-likelihood fit for the time-dependent analysis.  ``$\Bp\Bm$ ($\Bz\Bzb$) bkg'' represents background from charged (neutral) \B decays. Quoted uncertainties are statistical only.}
%\begin{tabular}{ lcc}
%\hline \hline \noalign{\smallskip}
%Species       & $\ 3\KS(\pip\pim)\ $     & $\ 2\KS(\pip\pim)\KS(\piz\piz)\ $\\ 
%\noalign{\smallskip}\hline\noalign{\smallskip}
%Signal        & $\rm 201\,^{+16}_{-15}$    & $\rm 62\,^{+13}_{-12}$ \\[0.5mm]
%Continuum     & $\rm 3086\,^{+56}_{-54} $  & $\rm 7086\,^{+85}_{-83} $ \\[0.5mm]
%$\Bp\Bm$ bkg  & $\rm -54\,^{+29}_{-24}$ &  $\rm 45\,^{+34}_{-30}$ \\[0.5mm]
%$\Bz\Bzb$ bkg & $\rm 9\,^{+31}_{-30}$ & $\rm 4\,^{+38}_{-29}$ \\[0.7mm] \hline \hline
%\end{tabular}             
%\end{center}
%\end{table}
%
The result is
\begin{eqnarray}
\nonumber {\cal S} &=& -0.94\,^{+0.24}_{-0.21} \pm 0.06, \\
\nonumber {\cal C} &=& -0.17 \pm 0.18 \pm 0.04, 
\end{eqnarray}
where the first quoted uncertainty is statistical and the second is systematic.
The correlation between ${\cal S}$ and ${\cal C}$ is $-0.16$.

\begin{figure}[htb]
\begin{center}
\includegraphics[width=7.0cm,keepaspectratio]{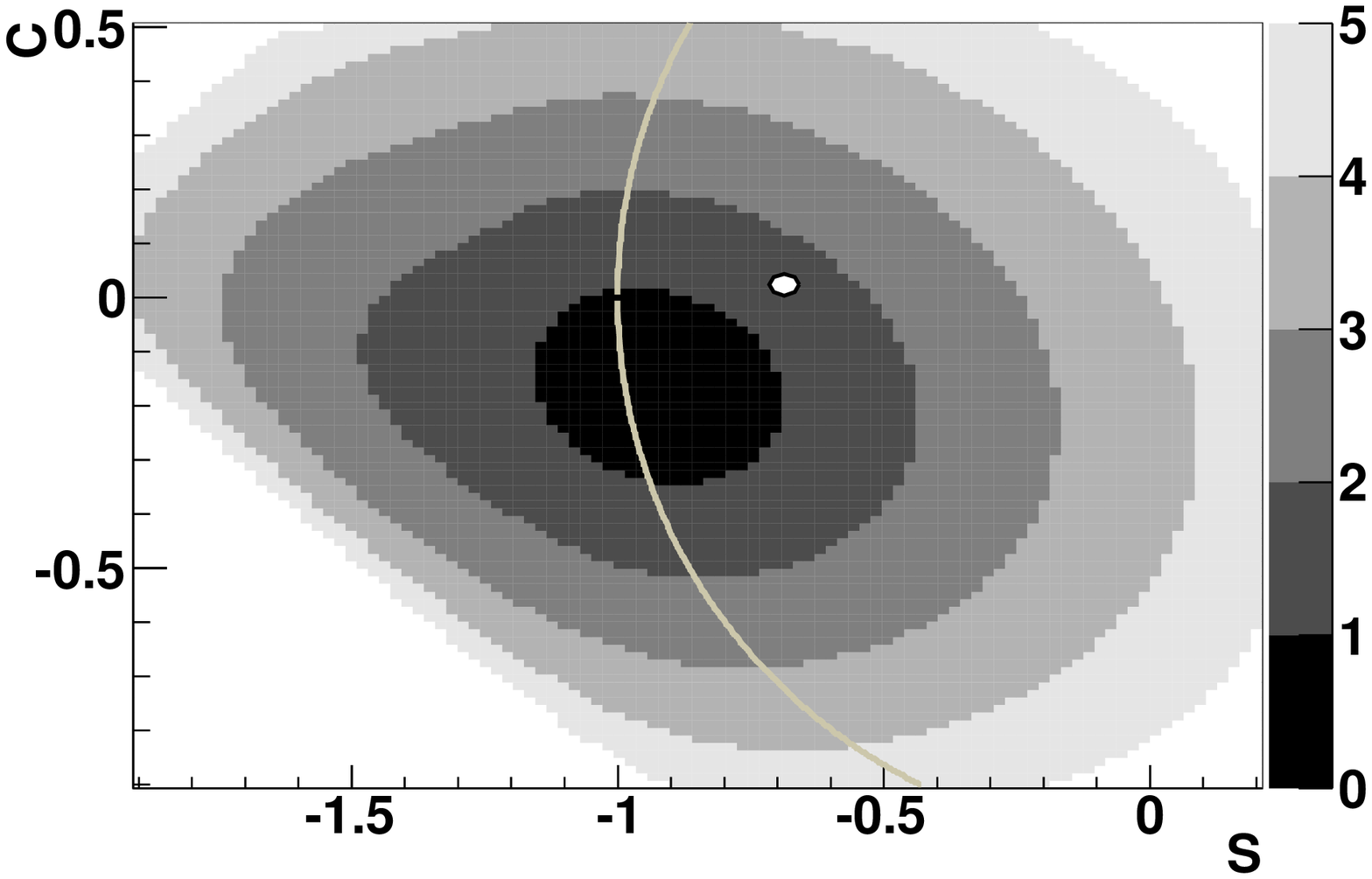}
\end{center}
\caption{Two-dimensional likelihood scan of $-2 \Delta \lnL$ as a function of ${\cal S}$ and ${\cal C}$, including systematic uncertainty, in $\Bz\to\KS\KS\KS$. %In the left-hand plot, red points marked with $\times$ correspond to the $\Bz\to 3\KS(\pip\pim)$ submode, blue points marked with $\circ$ to the $\Bz\to 2\KS(\pip\pim)\KS(\piz\piz)$ submode, and  black points marked with $\ast$ to the combined fit.
The gray scale is given in units of $\sqrt{-2 \Delta \lnL}$.
The result of the \babar\ analyses of $\Bz\to c \bar c K^{(*)}$ decays~\cite{Aubert:2009aw} is indicated as a white ellipse and the physical boundary (${\cal S}^2+{\cal C}^2 \leq 1$) is marked as a gray line. The scan appears to be trimmed on the lower left since the PDF becomes negative outside the physical region (i.e., the white region does not indicate that the scan flattens out at 5$\sigma$).}
\label{fig:likelihoodScans_TD}
\end{figure}

A two-dimensional statistical likelihood scan in ${\cal S}$ and ${\cal C}$, which is then convolved by the systematic uncertainties on the two parameters,
is shown in Fig.~\ref{fig:likelihoodScans_TD}.
We find that \CP\ conservation is excluded at $3.8$ standard deviations, and thus, for the first time, we measure an evidence of \CP\ violation in $\Bz\to\KS\KS\KS$ decays.
The difference between our result and that from $\Bz\to c \bar c K^{(*)}$ is less than $2$ standard deviations.
The scan also shows that the result is close to the physical boundary, given by the constraint ${\cal S}^2+{\cal C}^2 \leq 1$.

%%%%%%%%%%%%%%%%%%%%%%%%%%%%%%%%%%%%%%%%%%%%%%%%%%%%%%%%
\section{Dalitz-plot analysis of $\Bz \to \Kp \Km \KS$ decays}

Unlike $\Ksksks$, \kkks is not a \CP eigenstate, and therefore a Dalitz-plot analysis is necessary in order to disentangle \CP-even from \CP-odd states and measure $\betaeff$.
Because of the importance of understanding the Dalitz-plot structure in \bkkks, we
study the related modes \bkkkboth and \bkksks along with \bkkks.  
The mode \bkkkboth is valuable because it has the most
signal events by far of any $\bkkkgeneric$ mode.
Far fewer events are observed in $\bkksks$, but its
Dalitz-plot has a simplified spin-structure  due to the fact that the two $\KS$ mesons in the final state are forbidden (by Bose-Einstein statistics) to be in an odd angular momentum configuration.
A further benefit of a Dalitz-plot analysis is that it allows
both $\sin{2\betaeff}$ and  $\cos{2\betaeff}$ to be determined, through
the interference of odd and even partial waves, which eliminates a
trigonometric ambiguity between $\betaeff$ and $90^\circ - \betaeff$.

We describe the distribution of signal events 
in the Dalitz plot using an isobar approximation,
which models the total amplitude as
a coherent sum of amplitudes from individual decay channels (``isobars''). \KS\ mesons are reconstructed both in \Kspp and \Kszz.

\begin{figure}[htb]
\begin{center}
\includegraphics[width=7.5cm,keepaspectratio]{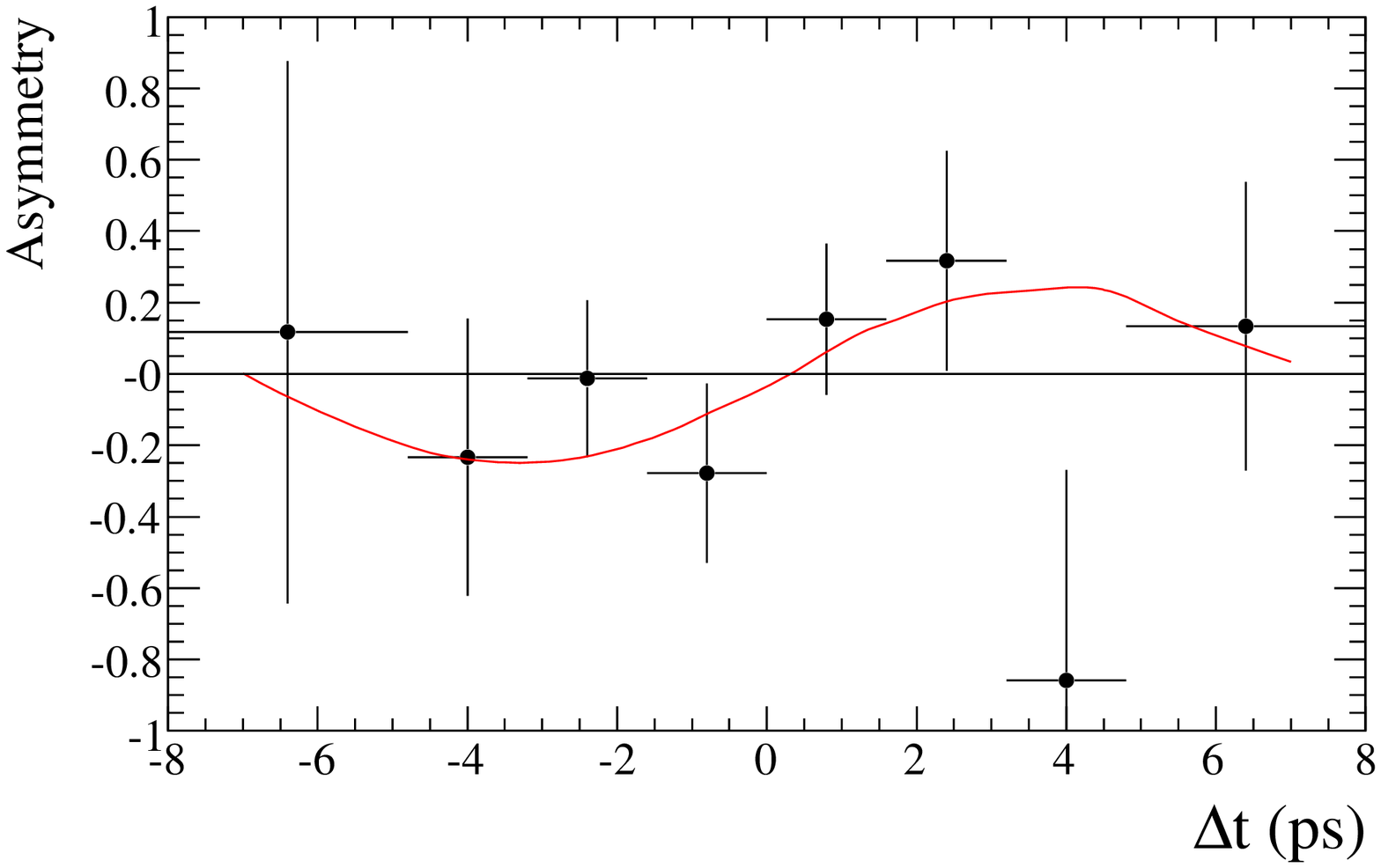}
\includegraphics[width=7.5cm,keepaspectratio]{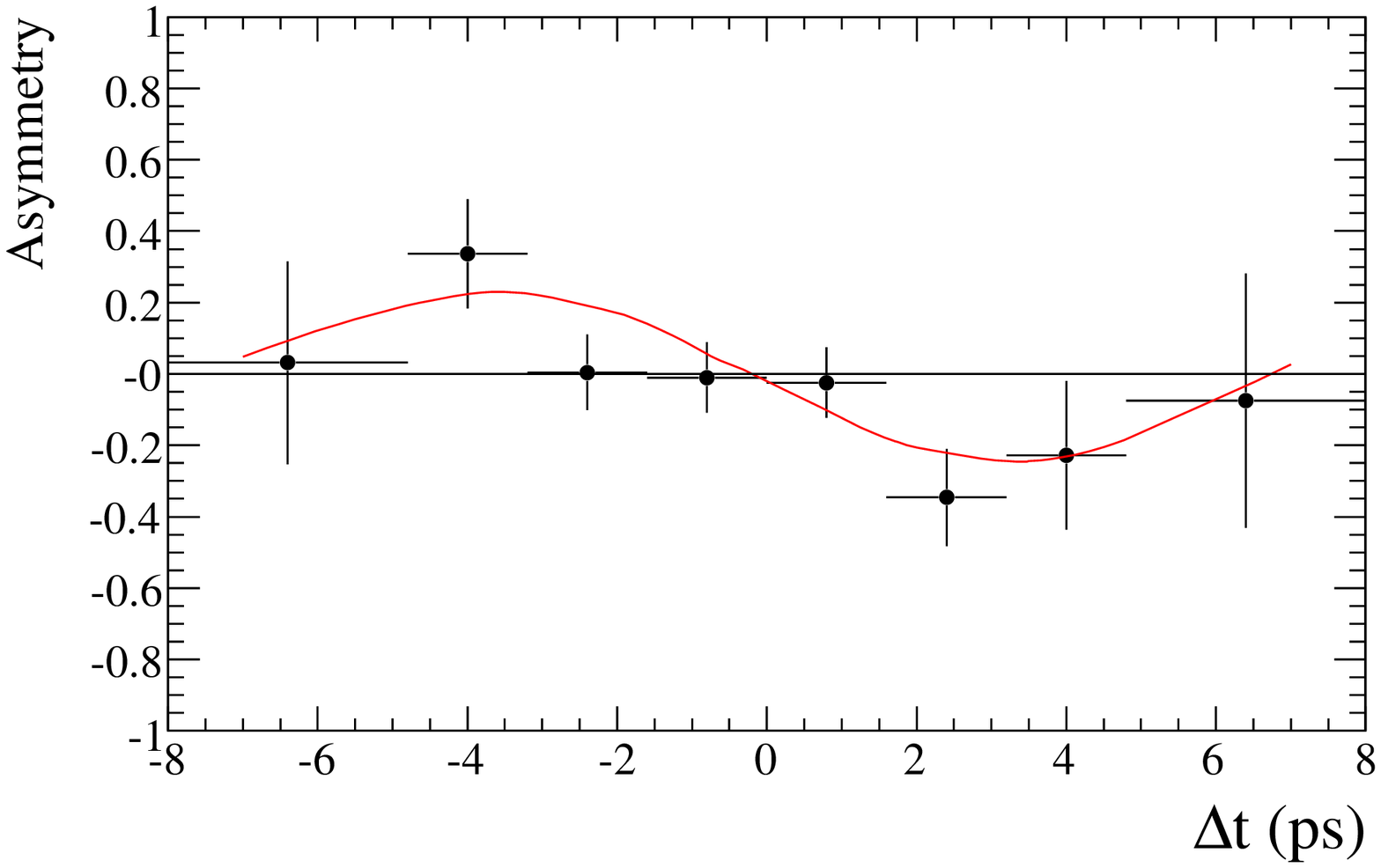}
\caption{\label{fig:bkkks_deltat}  
%Top:  The \deltat distributions for \bkkks (\Kspp) signal events, in the $\phiI$
%region $(1.01 <\mab<1.03 \gevcc)$ (left) and $\phiI$-excluded region (right).  
%\Bz (\Bzb) tagged events are
%shown as closed circles (open squares).  The fit model for \Bz (\Bzb) 
%tagged events is shown by a solid (dashed) line.  The data points are signal-weighted using the
%\splot method.
%Bottom: 
The asymmetry $(N_{\Bz} - N_{\Bzb})/(N_{\Bz} + N_{\Bzb})$ in \bkkks as a function of \deltat, 
in the $\phiI$ region (left) and $\phiI$-excluded region (right).  The points represent
signal-weighted data, and the line is the fit model.
}
\end{center}
\end{figure}

We vary three sets of $\betaeff$ and the \CP asymmetry, $\Acp$, values in the fit: one 
for the $\phiI$, another for the $\fI$, and a third that 
is shared by all the other charmless isobars in order to reduce the
number of fit parameters.  Note that this last set of isobars contains
both even-spin and odd-spin (P-wave) non-resonant terms. Figure~\ref{fig:bkkks_deltat} shows  the time-dependent asymmetry for signal-weighted events, both for
the $\phiI$ region $(1.01 <m_{KK}<1.03 \gevcc)$ and the $\phiI$-excluded region.
We find five solutions that are essentially degenerate in $\betaeff$,
and determine $\betaeff(\phi\KS) = (21\pm 6 \pm 2)^\circ$ from a likelihood scan.
This is the most accurate result for $\betaeff(\phi\KS)$ at present. 
Excluding the $\phiI\KS$ and $\fI\KS$ contributions, we  measure 
$\betaeff = (20.3\pm 4.3 \pm 1.2)^\circ$
for  the remaining $\bkkks$ decays, and 
exclude the trigonometric reflection $90^\circ - \betaeff$ at $4.8\sigma$,
including systematic uncertainties. All the results are in a good agreement with $\beta$, measured in $b \to c\bar{c}s$ decays.

%%%%%%%%%%%%%%%%%%%%%%%%%%%%%%%%%%%%%%%%%%%%%%%%%%%%%%%%
\section{Time-dependent \CP asymmetry of partially reconstructed \Bztodstdst\ decays}

Since \Bztodstdst is the decay of a scalar to two vector mesons,
the final state is a mixture of \CP eigenstates. The \CP-odd and \CP-even fractions, as well as the corresponding \CP asymmetries
have been previously measured from the angular analysis of completely reconstructed \Bztodstdst decays both by \babar\,\cite{Aubert:2009rx} and Belle\,\cite{Vervink:2009sy}.
Here, we report a recent measurement from \babar\ based on
the technique of partial reconstruction, which allows one to gain a
factor of $\simeq 5$ in the number of selected signal events 
with respect to the most recent \babar\ full reconstruction analysis in\,\cite{Aubert:2009rx}. 
This result is complementary to the latter measurement,
because the statistics used are largely independent of each other. 

In the partial reconstruction of a \Bztodstdst candidate, we fully reconstruct only one of the two \Dstarpm mesons in the decay chain \Dstar\to $D^0\pi$,
by identifying \Dz candidates in one of four final states: \ModeO, \ModeT, \ModeTh, \ModeF. Since the kinetic energy available in the decay \Dstar\to\Dz$\pi$ is small,
we combine one reconstructed \Dstarpm with an oppositely charged low-momentum pion,
assumed to originate from the decay of the unreconstructed \Dstarmp,
and evaluate the mass \mrec\ of the recoiling \Dz meson and its direction by using the
momenta of the two particles.
For signal events \mrec peaks at the nominal \Dz mass, while for background events no such peak is visible.
The direction of the recoiling \Dz meson helps to reduce the additional dilution in \B flavor tagging due to tagging tracks from the unreconstructed \Dz.

The analysis 
proceeds with a series of unbinned maximum-likelihood fits, performed simultaneously on the on- and
off-resonance data samples. The procedure can be logically divided in the following three steps: determination of the signal fraction and several PDF-shape parameters, determination of the tagging dilution due to wrong tag assignments, and finally the time-dependent fit to the data, fixing all parameter values obtained in the previous steps.

The final results for \calC and \calS, with their correlation coefficient $\rho$ are:
\begin{eqnarray}
\begin{array}{lll}
\calC & = &  +0.15\pm 0.09\pm 0.04 \quad\multirow{2}{*}{$\rho=0.0649$.}\\ 
\calS & = &  -0.34\pm 0.12\pm 0.05\nonumber
\end{array}
\label{eq:result-combined}
\end{eqnarray}
The measured values of \calS and \calC that we
obtain from data only represent a weighted average of the
\CP-even and \CP-odd wave function components. 
If penguin amplitudes can be neglected then $S_+=-S_-$, $C_+=-C_-$  
and the value of the \CP-even components \Sp and \Cp, 
which we are interested in, can be obtained using the relations:
\begin{eqnarray}
\begin{array}{lll}
\calC&=&C_+\\
\calS&=&S_+ \left(1-2 R_\perp\right),\nonumber
\end{array}
\label{eqn:S_C_and_Rperp-bis}
\end{eqnarray}
where the factor $(1-2R_\perp)$ represents the dilution introduced by
the \CP-odd component $R_\perp$ in the signal. 
To compute $S_+$
 we use the value measured by \babar\ of ($R_\perp=0.158\pm 0.029$)\,\cite{Aubert:2009rx}, 
where the uncertainty is the combined statistical and systematic.
To evaluate the related systematic uncertainty, we vary this value by $\pm 1\sigma$. 
We obtain
\begin{eqnarray}
\begin{array}{lll}
\Cp & = & +0.15  \pm 0.09 \pm 0.04 \\
\Sp & = & -0.49 \pm 0.18 \pm 0.07 \pm 0.04, \nonumber
\end{array}
\label{eq:SpCp}
\end{eqnarray}
where the uncertainties shown are statistical and systematic; the third uncertainty is
the contribution from the error on $R_\perp$ described above.
This result is compatible with previous
measurements from \babar\,\cite{Aubert:2009rx} and
Belle\,\cite{Vervink:2009sy} using fully reconstructed decays. It is approximately 20\% more accurate than the previous \babar\ measurement. 
This result well agrees with the standard model expectation of negligible
contributions to the decay amplitude from penguin diagrams and thence with $\Sp=-\sin
2\beta$.

%%%%%%%%%%%%%%%%%%%%%%%%%%%%%%%%%%%%%%%%%%%%%%%%%%%%%%%%%%%%%%%%%%%%%%%%%
%\begin{figure}[htb]
%\begin{center}
%\epsfig{file=KCaSc.eps,height=1.5in}
%\caption{Plan of the magnet used in the Mesmeric studies.}
%\label{fig:magnet}
%\end{center}
%\end{figure}
%%%%%%%%%%%%%%%%%%%%%%%%%%%%%%%%%%%%%%%%%%%%%%%%%%%%%%%%%%%%%%%%%%%%%%%%%%%

%%%%%%%%%%%%%%%%%%%%%%%%%%%%%%%%%%%%%%%%%%%%%%%%%%%%%%%%%%%%%%%%%%%%%%%%%
%\begin{table}[b]
%\begin{center}
%\begin{tabular}{l|ccc}  
%Patient &  Initial level($\mu$g/cc) &  w. Magnet &  
%w. Magnet and Sound \\ \hline
% Guglielmo B.  &   0.12     &     0.10      &     0.001  \\
% Ferrando di N. &  0.15     &     0.11      &  $< 0.0005$ \\ \hline
%\end{tabular}
%\caption{Blood cyanide levels for the two patients.}
%\label{tab:blood}
%\end{center}
%\end{table}
%%%%%%%%%%%%%%%%%%%%%%%%%%%%%%%%%%%%%%%%%%%%%%%%%%%%%%%%%%%%%%%%%%%%%%%%%%%

\def\Discussion{
\setlength{\parskip}{0.3cm}\setlength{\parindent}{0.0cm}
     \bigskip\bigskip      {\Large {\bf Discussion}} \bigskip}
\def\speaker#1{{\bf #1:}\ }
\def\endDiscussion{}

%\Discussion
%
%\speaker{D. Giovanni (University of Seville)}  My analysis indicates that the
%recovery of the two gentlemen is due simply to their embrace of the masculine
%principle and has nothing to do with magnetism at all.  Could you comment on 
%this?
%
%\speaker{Reggiano} Professor Giovanni has discussed this hypothesis in several
%forums, but, I do not believe there is anything in print.  I understand that
%he is spending his time in other pursuits.
%
%\speaker{D. Anna (University of Seville)}  In fact, my colleague Giovanni 
%has expressed opposite opinions on this question at various times, depending
%on the audience.  All of these testosterone-based theories are, of course,
%nonsense.
%
%\endDiscussion
 
\end{document}